\documentclass[10pt, conference, compsocconf]{IEEEtran}
\usepackage{fancyhdr}
\usepackage[pdftex]{graphicx}
\usepackage{amsfonts, amsmath, amssymb}
\usepackage{booktabs}
\usepackage{multirow}
\usepackage{graphicx, cite, color, setspace, url, verbatim, afterpage}
\usepackage{stfloats}
\usepackage{xspace}
\usepackage{paralist}
\usepackage{textcomp}

\usepackage{algorithm}
\usepackage[noend]{algpseudocode}

\usepackage{listings}

\newcommand{\figref}[1]{Figure~\ref{#1}}

\newcommand{\ignore}[1]{}

\def\naively{na\"\i vely}
\def\ie{{\it i.e.}}
\def\eg{{\it e.g.}}

\begin{document}

 \author{Rohit Zambre$^{\star}$, Lars Bergstrom$^{\dagger}$,  Laleh Aghababaie Beni$^{\ast}$, Aparna Chandramowlishwaran$^{\star}$
 \vspace{0.1in}\\
 {\em  $^{\star}$EECS, University of California, Irvine, CA}\\
 {\em  $^{\ast}$ICS, University of California, Irvine, CA} \\
 {\em  $^{\dagger}$Mozilla Research, USA}\\
 }

\title{Parallel Performance-Energy Predictive Modeling of Browsers: Case Study of Servo}

\maketitle

\begin{abstract}
	
	Mozilla Research is developing Servo, a parallel web browser engine, to exploit the benefits of parallelism and concurrency in the web rendering pipeline. Parallelization results in improved performance for \emph{pinterest.com} but not for \emph{google.com}. This is because the workload of a browser is dependent on the web page it is rendering. In many cases, the overhead of creating, deleting, and coordinating parallel work outweighs any of its benefits. In this paper, we model the relationship between web page primitives and a web browser's parallel performance using supervised learning. We discover a feature space that is representative of the parallelism available in a web page and characterize it using seven key features. Additionally, we consider energy usage trade-offs for different levels of performance improvements using automated labeling algorithms. Such a model allows us to predict the degree of parallelism available in a web page and decide whether or not to render a web page in parallel. This modeling is critical for improving the browser's performance and minimizing its energy usage. We evaluate our model by using Servo's layout stage as a case study. Experiments on a quad-core Intel Ivy Bridge (i7-3615QM) laptop show that we can improve performance and energy usage by up to 94.52\% and 46.32\% respectively on the 535 web pages considered in this study. Looking forward, we identify opportunities to apply this model to other stages of a browser's architecture as well as other performance- and energy-critical devices. 
	
\end{abstract}


%

\section{Introduction}
\label{sec:introduction}

\begin{figure*}[htbp]
	\begin{center}
		\includegraphics[width=\textwidth]{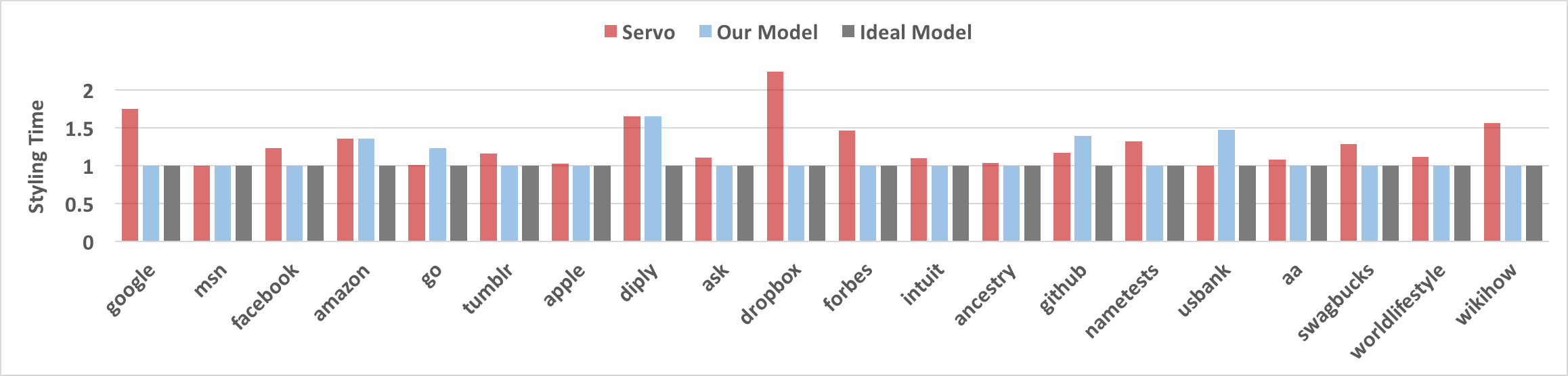}
	\end{center}
	\caption{Normalized styling times of Servo, Our Model, and Ideal Model.}
	\label{fig:servoVsModel}
\end{figure*}%

For any particular browser, a heavier page takes longer to load than a lighter one~\cite{lightheavypage}. The workload of a web browser is dependent on the web page it is rendering. Additionally, with the meteoric rise of the Web's popularity since the 1990s, web pages have become increasingly dynamic and graphically rich---their computational complexity is increasing. Hence, the page load times of web browsers have become a growing concern, especially when the user experience affects sales. A two-second delay in page load time during a transaction can result in abandonment rates of up to 87\%~\cite{twoseconddelay}. Further challenging matters, an optimization that works well for one page may not work for another~\cite{nejati2016depth}.

The concern of slow page load times is even more acute on mobile devices. Under the same wireless network, mobile devices load pages $3\times$ slower than desktops, often taking more than 10 seconds~\cite{butkiewicz2015klotski}. As a result, mobile developers deploy their applications using low-level native frameworks (\eg{} Android, iOS, etc. applications) instead of the high-level browser, which is usually the case for laptop developers. However, these applications are hard to port to phones, tablets, and smart TVs, requiring the development and maintenance of a separate application for each platform. With a faster browser, the universal Web will become more viable for all platforms.

The web browser was not originally designed to handle the increased workload in today's web pages while still delivering a responsive, flicker-free experience for users. Neither was its core architecture designed to take advantage of the multi-core parallelism available in today's processors. An obvious way to solve the slow web browser problem is to build a parallel browser. Mozilla Research's Servo~\cite{SERVO} is a new web browser engine designed to improve both memory safety, through its use of the Rust~\cite{RUST} programming language, and responsiveness, by increasing concurrency, with the goal of enabling parallelism in \emph{all} parts of the web rendering pipeline.

Currently, Servo (see Section ~\ref{sec:servo}) parallelizes its tasks for all web pages without considering their characteristics. However, if we \naively{} attempt to parallelize web rendering tasks for all content, we will incur overheads from the use of excessive number of threads per web page. More importantly, we may also penalize very small workloads by increasing power usage or by delaying completion of tasks due to the overhead of coordinating parallel work. Thus, the challenge is to ensure fast and efficient page load times while preventing slowdowns caused by parallel overhead. We tackle this challenge by modeling, using accurate labels and supervised learning, the relationship between web page characteristics and the parallel performance of a web rendering engine and its energy usage within the \emph{complete} execution of a browser. In this paper, we work with Servo since it is currently the only publicly available parallel browser. However, our modeling approach can easily extend to any parallel browser on any platform since our feature space is \emph{blind} to the implementation of a web rendering engine.

Precisely, we model with seven web page features that represent the amount of parallelism available in the page. These features are oblivious to the implementation of a rendering engine. We correlate these features to the parallel performance in two stages of a parallel web rendering engine. The first stage, \emph{styling}, is the process in which the engine determines the CSS styles that apply to the various HTML elements in a page. The second stage analyzed in this work is \emph{layout}. During \emph{layout}, the engine determines the final geometric positions of all of the HTML elements. We choose these two stages since they consume a significant portion of overall rendering time, especially for modern dynamic web pages. Internet Explorer and Safari spend 40-70\% of their web page processing time, on an average, in the visual layout of the page~\cite{meyerovich2010fast}.

We evaluate our model for Servo on a quad-core Intel Ivy Bridge using off-the-shelf supervised learning methods on 535 web pages. Even with a large class-imbalance in our data set, we demonstrate strong accuracies, approaching 88\%. \figref{fig:servoVsModel} depicts the styling times taken by Servo and our model against an optimal model (to which times are normalized) for the top 20 web pages in the Alexa Top 500~\cite{alexa-top-500} list. \emph{Ideal Model} represents the best time that is achieved using either 1, 2 or 4 threads (see Section ~\ref{sec:setup} for our experiment's configuration). \emph{Our Model} represents the time taken using the number of threads suggested by our proposed model. \emph{Servo} represents the time taken by 4 threads, the default number of threads that the Servo browser engine, unlike our model, spawns for the styling and layout stages on a quad-core processor. \emph{Our Model} performs as well as the \emph{Ideal Model} in most cases.

We make three main contributions:

\begin{enumerate}
\item \textbf{Workload characterization --} The workload of a browser is dependent on the web page. We study and analyze the Document Object Model (DOM)~\cite{domdef} tree characteristics of a web page and use them to characterize the parallel workload of the rendering engine (see Section ~\ref{sec:char}).
  
\item \textbf{Performance-energy labeling of web pages --} Considering performance speedups and energy usage ``greenups,"~\cite{choi2013} we label web pages into different categories using three cost models. To do so, we propose automated labeling algorithms for each cost model (see Section ~\ref{sec:classification}).

\item \textbf{Performance-energy modeling and prediction --} Using supervised learning, we construct, train, and evaluate our proposed statistical inference models that capture the relationship between web page characteristics and the rendering engine's performance and energy usage. Given the features of a web page, we use the model to answer two fundamental questions: (a) should we parallelize the styling and layout tasks for this web page? If so, (b) what is the degree of available parallelism? (see Section ~\ref{sec:perf})

\end{enumerate}

\section{Web Page Characterization}
\label{sec:char}

\begin{figure}[htbp]
	\centering	
	\begin{minipage}[t]{0.27\textwidth}\centering\includegraphics[width=\textwidth]{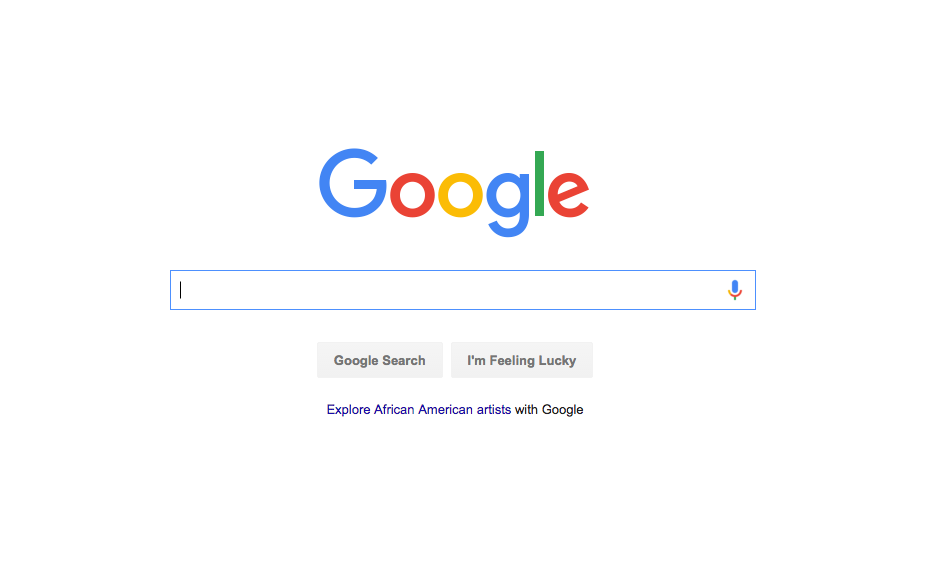}\end{minipage}
	\begin{minipage}[t]{0.21\textwidth}\centering\includegraphics[width=\textwidth]{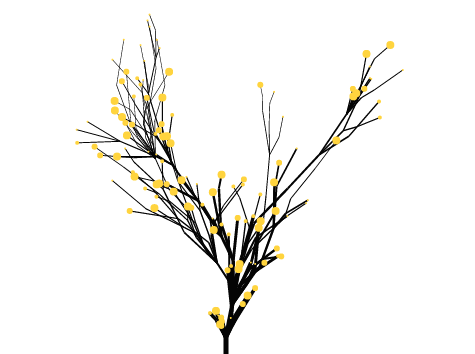}\end{minipage}\\
	\vspace{10 pt}
	(a) \texttt{google.com}\\
	\vspace{10 pt}
	\begin{minipage}[t]{0.27\textwidth}\centering\includegraphics[width=\textwidth]{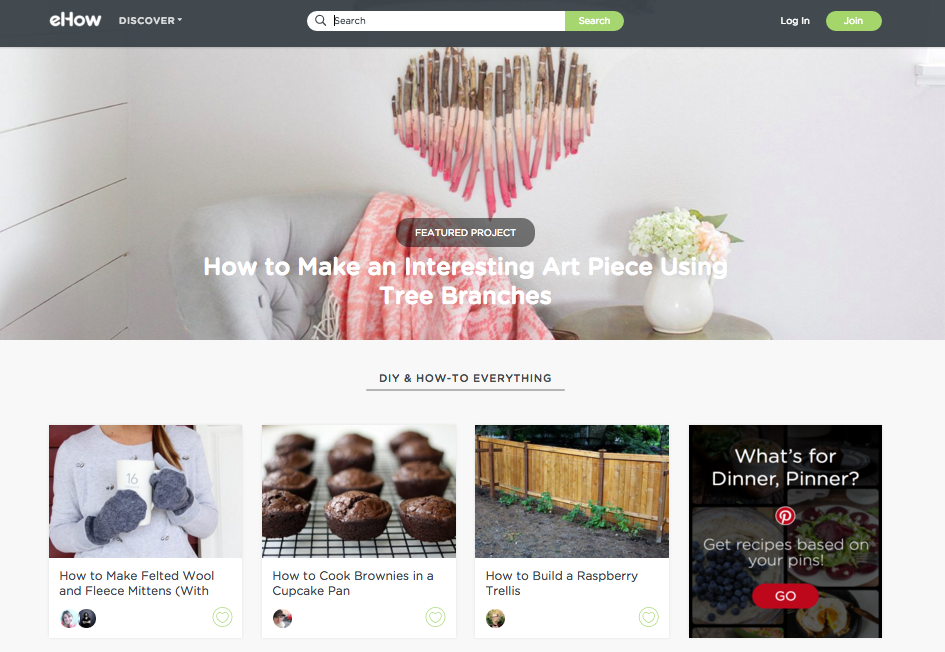}\end{minipage}
	\begin{minipage}[t]{0.21\textwidth}\centering\includegraphics[width=\textwidth]{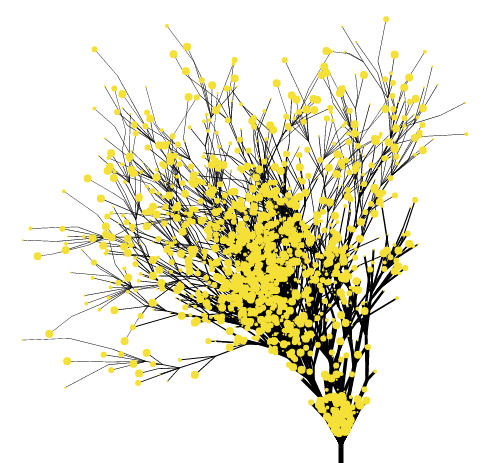}\end{minipage}\\
	\vspace{10 pt}
	(b) \texttt{ehow.com}\\
	\caption{Contrasting types of web pages (left) and the visualization of their DOM trees (right).}
	\label{fig:contrasting-examples}
\end{figure}

A wide variety of web pages exists in today's World Wide Web. Either a web page can contain minimal content with little to no images or text, or it can include a wide variety of multimedia content including images and videos. The left column of \figref{fig:contrasting-examples} depicts the web pages of \texttt{google.com} and \texttt{ehow.com}, two contrasting instances that exemplify the variety of web pages that one comes across on a daily basis.

\begin{figure*}[htbp]
	\centering
	\begin{minipage}[t]{0.24\textwidth}\centering\includegraphics[width=\textwidth]{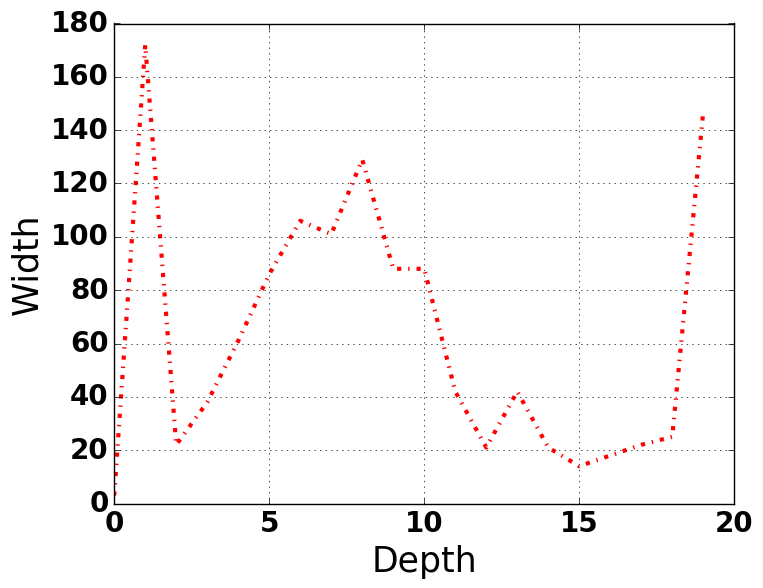}\par(a) \texttt{airbnb.com} \end{minipage}
	\begin{minipage}[t]{0.24\textwidth}\centering\includegraphics[width=\textwidth]{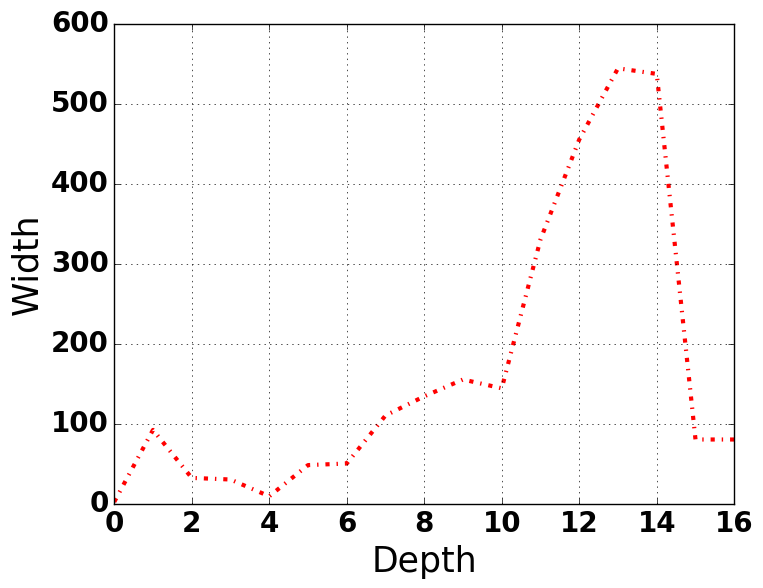}\par(b) \texttt{samsclub.com} \end{minipage}
	\begin{minipage}[t]{0.24\textwidth}\centering\includegraphics[width=\linewidth]{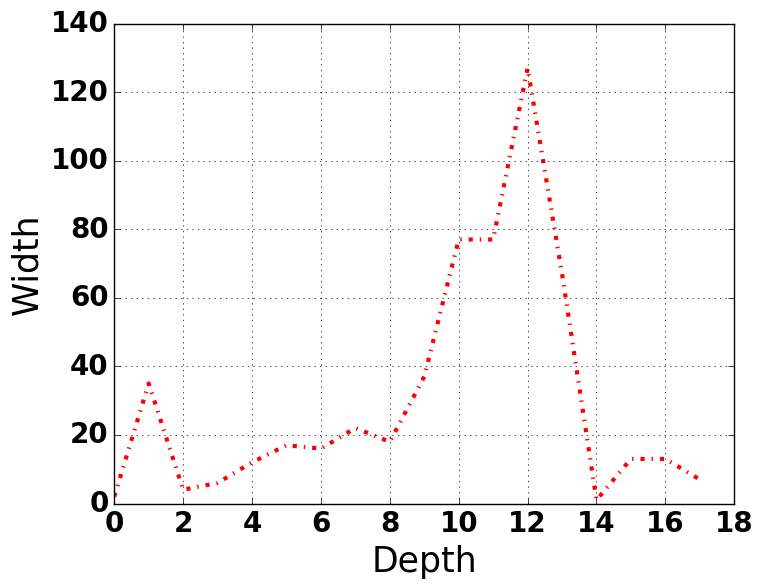}\par(c) \texttt{westlake.com} \end{minipage}
	\begin{minipage}[t]{0.24\textwidth}\centering\includegraphics[width=\linewidth]{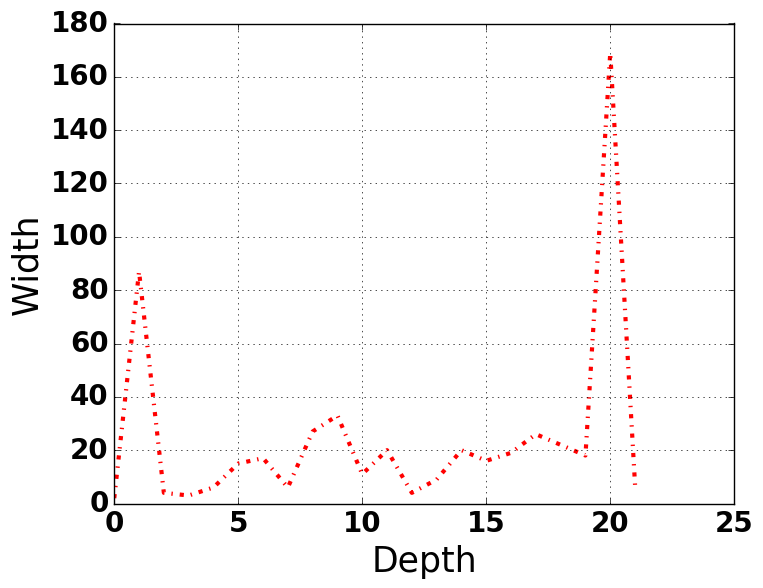}\par(d) \texttt{facebook.com} \end{minipage}
	\vspace{1em}
	\caption{Width Vs. Depth graphs of the DOM trees of different web pages (note that the scales of the axes are different).}
	\label{fig:dom-trees}
\end{figure*}

The right column of \figref{fig:contrasting-examples} portrays a visual representation (created using Treeify~\cite{treeify}) of the DOM tree of the corresponding web pages. The DOM tree is an object representation of a web page's HTML markup. Qualitatively, \figref{fig:contrasting-examples} shows that simple web pages, like \texttt{google.com}, have relatively small DOM trees, low number of leaves, and are not as wide or as deep. On the other hand, complex web pages, such as \texttt{ehow.com}, have relatively big trees, a high number of leaves, and are much wider and deeper.

Browser optimizations are primarily applied to the style application stage since it is the most CPU-intensive of all stages~\cite{html}. It is during this step that the DOM tree is traversed extensively to compute the styles for each element on the page. Naturally, a parallel browser would then optimize this stage using parallel tree traversals~\cite{meyerovich2010fast}. Hence, we identify DOM tree features that correlate strongly with the performance of these parallel tree traversals. Any amount of parallel speedup or slowdown would depend on the structure of the DOM tree. We intuitively choose the following set of nine characteristics to capture the properties of a web page and its DOM tree:

\begin{enumerate}[1.]
	\item Total number of nodes in the DOM tree (\textbf{DOM-size})
	\item Total number of attributes in the HTML tags used to describe the web page (\textbf{attribute-count})
	\item Size of the web page's HTML in bytes (\textbf{web-page-size})
	\item Number of levels in the DOM tree (\textbf{tree-depth})
	\item Number of leaves in the tree (\textbf{number-of-leaves})
	\item Average number of nodes at each level of the tree (\textbf{avg-tree-width})
	\item Maximum number of nodes at a level of the tree (\textbf{max-tree-width})
	\item Ratio of max-tree-width to average-tree-width (\textbf{max-avg-width-ratio})
	\item Average number of nodes per level of the tree (\textbf{avg-work-per-level})
\end{enumerate}

Our intuition is that large\footnote{The values of the features lie on a continuous spectrum and so, we cannot assign discrete definition to descriptors such as ``big," ``large," ``small," ``wide," ``narrow," etc.} and wide trees observe higher speedups than small and narrow trees in parallel tree traversals (captured by \textbf{DOM-size} and \textbf{avg-tree-width}). In consecutive top-down and bottom-up parallel traversals (see Section~\ref{sec:servo}), trees with a large number of leaves observe faster total traversal completion time (captured by \textbf{number-of-leaves}). Even amongst wide trees, those that don't have abrupt changes in tree-width, or are less deep, observe faster parallel traversals (captured by \textbf{tree-depth}, \textbf{max-avg-width-ratio}). \textbf{DOM-size} captures the total amount of work while \textbf{avg-work-per-level} ($=\textbf{DOM-size}/\textbf{tree-depth}$) captures the average parallel work on the web page. \textbf{attribute-count}, and \textbf{web-page-size} capture the general HTML information about a web page. Although we initially identify nine features, we only choose seven for modeling based on the results of the statistical correlation of these characteristics to Servo's parallel performance (see Section~\ref{sec:perf}).


To quantitatively analyze DOM trees, we plot the width of the trees at each of their depth levels. In \figref{fig:dom-trees}, we do so for \texttt{airbnb.com}, \texttt{samsclub.com}, \texttt{westlake.com}, and \texttt{facebook.com}. Using these figures, we relate our intuition to observed data. \texttt{airbnb.com} and \texttt{samsclub.com} are examples of DOM tree structures that represent ``good" levels of available parallelism. The \textbf{DOM-size} of \texttt{samsclub.com} is $2833$ and hence, sufficient work is available. The \textbf{DOM-size} of \texttt{airbnb.com} is $1247$ which is much smaller than that of \texttt{samsclub.com}. However, the DOM tree of \texttt{airbnb.com} has a high \textbf{avg-tree-width} of $62.3$, a characteristic that favors parallelism. Our performance experiments show that \texttt{samsclub.com} achieves $1.48\times$ speedup with 2 threads and $2.12\times$ speedup with 4 threads. \texttt{airbnb.com} achieves speedups of $1.2\times$ and $1.43\times$ with 2 and 4 threads respectively, which, although significant, are not as high as those of \texttt{samsclub.com} due to the lesser amount of available work. The DOM trees of \texttt{westlake.com} and \texttt{facebook.com} exemplify tree structures that represent ``bad" candidates for parallelism. These trees have large widths only for a small number of depth levels. Hence, the \textbf{avg-tree-width}s of these trees are low: $30.6$ and $24.6$ for \texttt{westlake.com} and \texttt{facebook.com} respectively. These trees don't have enough amount of work to keep multiple threads occupied. \texttt{westlake.com} shows slowdowns of $0.94\times$ and $0.74\times$ with 2 and 4 threads respectively. Similarly, \texttt{facebook.com} demonstrates slowdowns of $0.86\times$and $0.81\times$ with 2 and 4 threads respectively.
\section{Servo Overview}
\label{sec:servo}


Servo~\cite{SERVO} is a web browser engine that is being designed and developed by Mozilla Research. The goal of the Servo project is to create a browser architecture that employs inter- and intra-task parallelism while eliminating common sources of bugs and security vulnerabilities associated with incorrect memory management and data races. C++ is poorly suited to prevent these problems.

\begin{figure}[htbp]
	\begin{center}
		\includegraphics[width=0.49\textwidth]{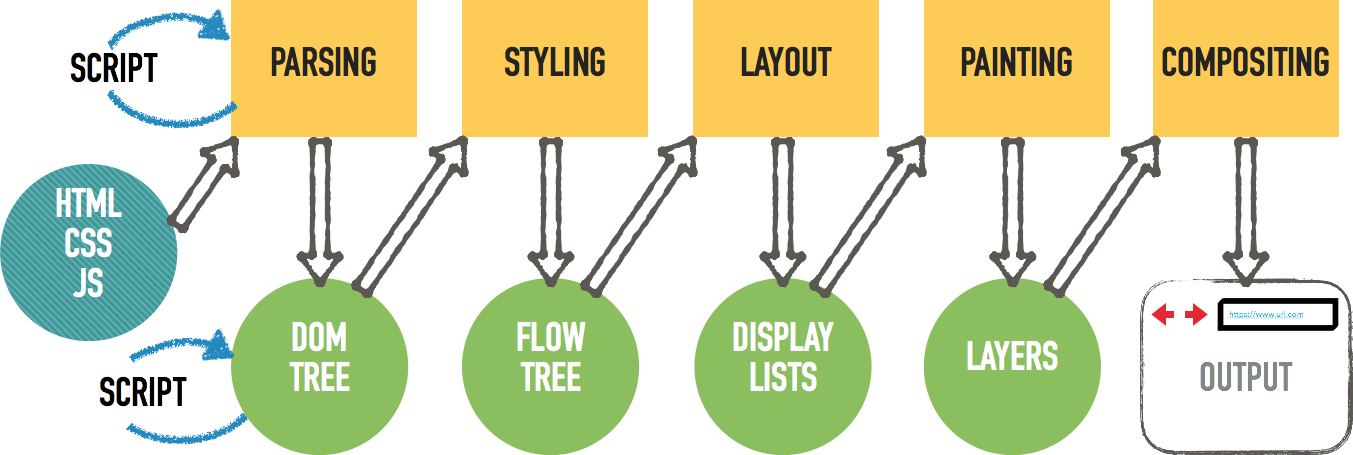}
	\end{center}%
	\vspace{-1em}
	\caption{Processing stages and intermediate representations in a browser engine. The circles represent data structures while the squares represent tasks.}
	\label{fig:browser}
	\vspace{-1em}
\end{figure}%

Servo is written in Rust~\cite{RUST}, a new language designed by Mozilla Research  specifically with Servo's requirements in mind. Rust provides a task-parallel infrastructure and a strong type system that enforces memory safety and data-race freedom. The Servo project is creating both a full web browser, through the use of the purely HTML-based user interface BrowserHtml~\cite{browserhtml}, and a solid embeddable web rendering engine. Although Servo was originally a research project, it was implemented with the goal of production-quality code and is in the process of shipping several of its components to the Firefox browser.

The processing steps used by all browsers are very similar, as many parts of the interpretation of web content are defined by the standards from the World Wide Web Consortium (W3C) and the Web Hypertext Application Technology Working Group (WHATWG).
As such, the steps Servo uses in \figref{fig:browser} should be unsurprising to those familiar with the implementation of other modern browsers~\cite{html}.
Due to space constraints, we describe only the phases relevant to this paper in detail.
A more detailed description of these stages is available in ~\cite{servo-icse}, which, orthogonal to this work, outlines the language features of Rust in the context of Servo.

The first step in loading a site is retrieving and \emph{parsing} the HTML and CSS files. The HTML translates into a DOM tree and the CSS loads into style structures.
Each node of the tree corresponds to an HTML element in the markup. The CSS style structures are used in \emph{styling}. JavaScript may execute twice, during parsing and after page-load during user-interactivity when it can modify the DOM tree.

\subsection{Styling}

After constructing the DOM tree, Servo attaches styling information in the style structures to this tree. In this process, it builds another tree called the \emph{flow tree} which describes the layout of the DOM elements on the page in the correct order. However, the flow tree and the DOM tree don't hold a one-to-one relation unlike that of the HTML markup and the DOM tree. For example, when a list item is styled to have an associated bullet, the bullet itself will be represented by a separate node in the flow tree even though it is not part of the DOM tree.

\begin{figure}[htbp]
	\begin{center}
		\includegraphics[width=0.49\textwidth]{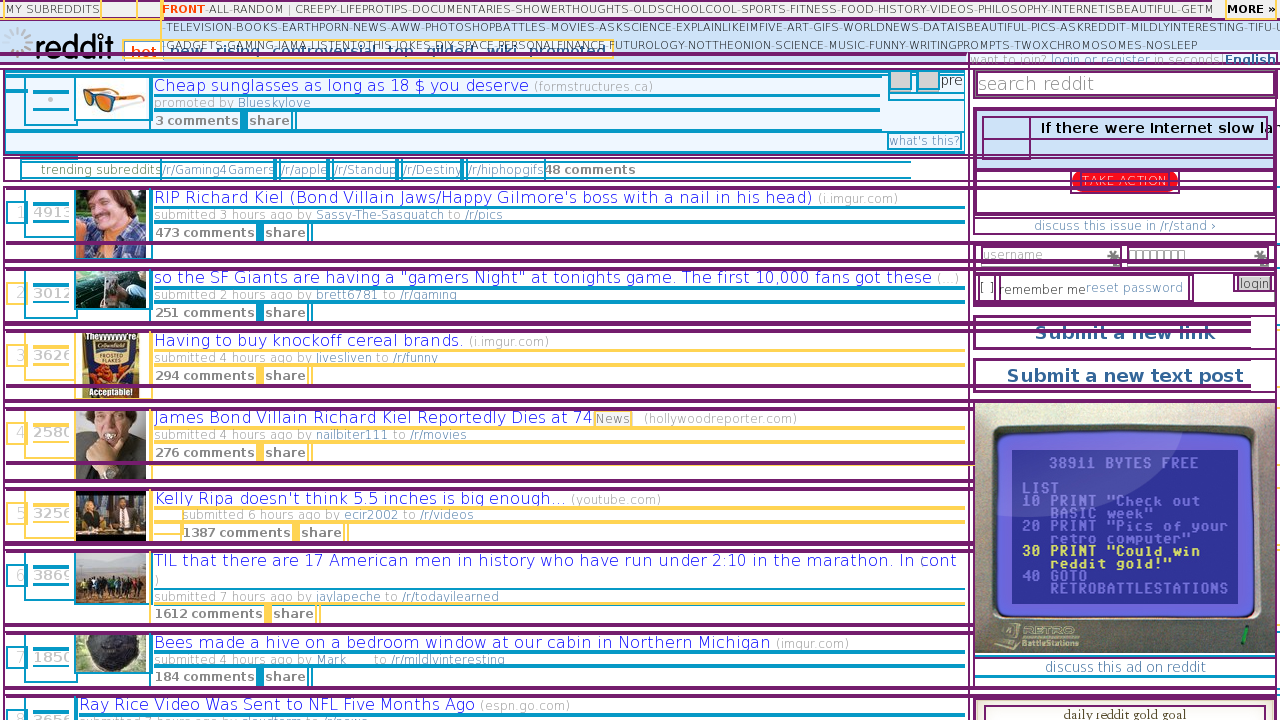}
	\end{center}%
	\vspace{-1em}
	\caption{Parallel layout on \texttt{reddit.com}.
		Different colors indicate that layout was performed by a different thread.}
	\label{fig:parlayout}
	\vspace{-1em}
\end{figure}%

This stage contains the first step that is the subject of analysis in this paper. Servo executes styling using parallel tree traversals, an approach similar to the one employed by Meyerovich et al.~\cite{meyerovich2010fast}. Conceptually, the first half of this step is a trivially parallel process---each DOM tree node's style can be determined at the same time as any other node's. However, to prevent massive memory growth, Servo shares the concrete style structures that are associated with multiple nodes, requiring communication between parallel threads. The second half of this step is the construction of the flow tree.

\subsection{Layout}

The flow tree is then processed to determine the final geometric positions of all the elements first and then to produce a set of \emph{display list} items.

Determining these positions is the second step that is analyzed in this paper. In cases where no HTML elements prevent simultaneous evaluation (\eg{} floated elements and mixed-direction writing modes), Servo performs consecutive top-down and bottom-up parallel tree traversals to determine the final positions of elements; the height of a parent node is dependent on the cumulative height of its children, and the widths of the children are reliant on the width of the parent. These traversals execute incrementally and hence multiple individual passes occur before the end of a page-load. \figref{fig:parlayout} shows one parallel execution with four cores rendering \texttt{reddit.com}. 

After final positions of the elements are computed, the engine constructs display list items. These list items are the actual graphical elements, text runs, etc. in their final on-screen positions. The order in which to display these items is well-defined by the CSS standard~\cite{w3}.

%
%
Finally, the to-be-displayed elements are \emph{painted} into memory buffers or directly onto graphic-surfaces (\emph{compositing}). Servo may paint each of these buffers in parallel.

In the rest of the paper, we will use the term \textbf{Overall Layout} to refer to the \emph{Styling} and \emph{Layout} stages together and \textbf{Primary Layout} to refer to the \emph{Layout} stage alone. Also, unless specified otherwise, we refer to \emph{total} times of all the incremental passes performed for each stage. The sequential baseline performance of Servo's \emph{Overall Layout} is nearly $2\times$ faster than Firefox's (Table 1 of~\cite{servo-icse}). This speedup stems from the use of Rust instead of C++ as more optimization opportunities exist in Rust. Additionally, the parallel implementation of \emph{Overall Layout} has been optimized \ie{} we observe high speedups on some websites: $3.2\times$ and $2.5\times$ with 4 threads on \texttt{humana.com} and \texttt{kohls.com} respectively. In some cases, 2 threads perform better than 4 threads: \texttt{walgreens.com} achieves $1.43\times$ speedup with 2 threads and $1.16\times$ with 4 threads. We attribute such slowdowns to the parallel overhead of synchronization, which we aim to mitigate with our modeling. Currently, Servo uses a work-stealing scheduler to schedule the threads that are spawned (once) to perform the parallel tree traversals in \emph{Overall Layout}.
\section{Experimental Setup}
\label{sec:setup}

The Web is extremely flexible and dynamic. Constantly changing network conditions can add a significant amount of variability in any testing that involves acquiring data directly from the Internet. Further, due to advertising networks, A/B testing, and rapidly changing site content, even consecutive requests can have significantly different workloads.

Hence, to achieve repeatable and reliable performance results with Servo, we use Google's Web Page Replay~\cite{WPR} (WPR) to \emph{record} and \emph{replay} the HTTP content required for our tests. At a high level, WPR establishes local DNS and HTTP servers; the DNS server points to the local HTTP server. During \emph{record}, the HTTP server acquires the requested content from the Internet and saves it to a local archive while serving the requesting client. During \emph{replay}, the HTTP server serves all requests from the recorded archive. A 404 is served for any content that is not within the archive. We used WPR to \emph{record} the web pages in our sample set first and then \emph{replay} them during the experiments. The \emph{replay} mode guarantees that Servo receives the same content every time it requests a particular web page. For our testing platform, we use a quad-core Intel i7-3615QM running OS X 10.9 with 8GB of RAM.

We collect our sample dataset from two sources: (1) Alexa Top 500~\cite{alexa-top-500} web pages in the United States during January 2016, and (2) 2012 Fortune 1000~\cite{2012-fortune-1000}. We initially started with $1000$ web pages but these contained domain names that were either outdated or corresponded to server names and not actual web pages (\eg{} \texttt{blogspot.com}, \texttt{t.co}). Also, some web pages caused runtime errors when Servo tried to load them since it is a work in progress. After filtering out all such web pages, we have a working set of $535$ web pages.

For our performance testing, we use Servo's internal profiling tool that spits out a CSV file containing user times of the \emph{Styling} and \emph{Primary Layout} stages. For our energy testing, we use Apple's powermetrics~\cite{powermetrics} to capture processor power usage. In both the energy and timing experiments, we call Servo to open a web page and terminate it as soon as the \emph{page load} is complete. Across all browsers, \emph{page load} entails fully loading and parsing all resources, reflecting any changes through re-styling and layout. Servo goes a bit further in the automation harness and will also wait until any changes to the display list have been rendered into graphics buffers and until the in-view graphics buffers composite to the final surface. In the energy experiments, we allowed a sleep time of $10$ seconds before each run of Servo to prevent incorrect power measurements due to continuous processor usage.

Servo makes it possible to specify the number of threads to spawn in the \emph{Overall Layout} stage. Given that our platform is a quad-core, we used $1$, $2$, and $4$ as the number of threads for each web page. Since Servo is still under development, we observe non-repeatable behavior. To account for repeatability, we run $5$ trials with each thread number for each web page for both the performance and energy experiments. With 5 trials, the medians of the \emph{median absolute deviations} (MAD) of all 1-, 2- and 4-thread executions are low: $6.76$, $7.46$, and $7.49$ respectively. MAD is a robust measure of the variability of a data sample~\cite{leys2013detecting}.

\section{Performance and Energy Automated Labeling}
\label{sec:classification}

\begin{figure*}[htbp]
	\begin{center}
		\includegraphics[width=\textwidth]{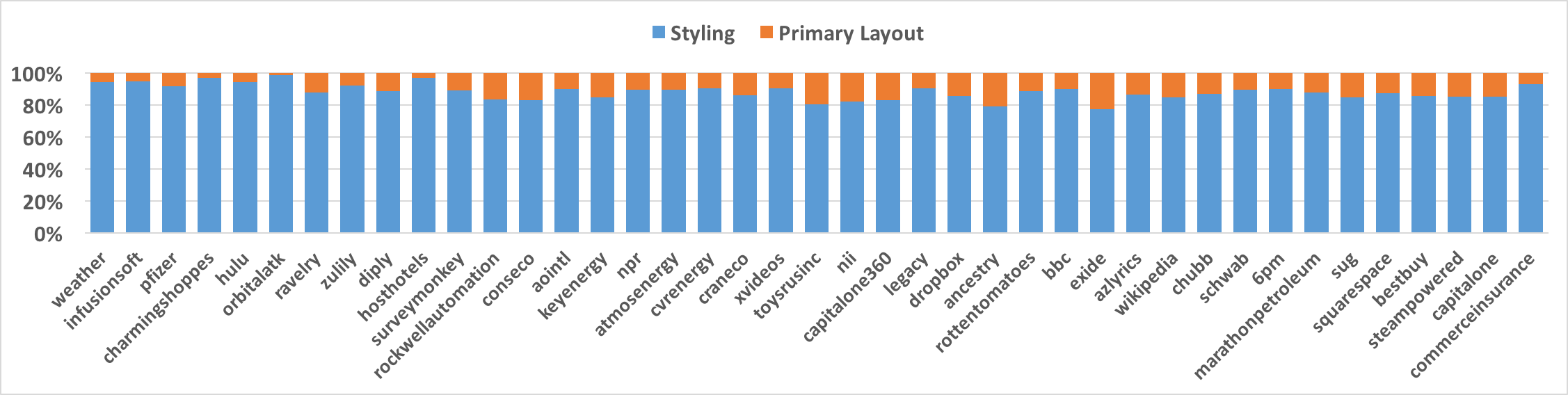}
	\end{center}%
	\vspace{-1em}
	\caption{Styling and Primary Layout time division in Overall Layout for a random sample of $40$ web pages}
	\label{fig:percentLayout}
\end{figure*}

In this section, we propose tunable, automated labeling algorithms that can be used with any web browser on any testing platform to classify web pages into different categories. Automated labeling eliminates the need for a domain expert to manually and accurately label data. Given the labels, a predictive model can be trained using supervised learning methods. For labeling, we consider three cost models:

\begin{enumerate}[1.]
	\item \textbf{Performance --} Labels depend only on performance improvements from parallelization.
	\item \textbf{Energy --}  Labels depend only on energy usage increases from parallelization.
	\item \textbf{Performance and Energy --} Labels depend on both performance improvements and energy usage increases from parallelization. 
\end{enumerate}

Although we collected user times for both the \emph{Styling} and \emph{Primary Layout} stages on Servo, we focus our comparisons on the former for the following reasons -- (1) The \emph{Styling} stage works on the DOM tree while the \emph{Primary Layout} stage works on the flow tree. Since we characterize web pages using the DOM tree, we expect \emph{Styling} performance to correlate strongly to the tree characteristics. (2) The medians of total \emph{Styling} time and total \emph{Primary Layout} time as percentages of the \emph{Overall Layout} time (single-thread execution) are $67.19\%$ and $7.83\%$. Clearly, \emph{Styling} time primarily defines \emph{Overall Layout} time. ~\figref{fig:percentLayout} shows these percentages for a sample of $40$ randomly selected web pages.


Another interesting observation is that an individual serial pass of \emph{Primary Layout} ranges between $1$ and $55$ ms. This range is much smaller than the $1$ to $320$ ms range of an individual serial \emph{Styling} pass. Hence, parallelizing tree traversals for the \emph{Primary Layout} stage will most likely result in poorer performance due to thread communication and scheduling overheads. Our results validate this analysis. On average, the time taken by an individual parallel pass for \emph{Primary Layout} is $3.92$ ms, $7.01$ ms, and $9.82$ ms with 1, 2, and 4 threads respectively. We also observe an increase in the average total times for \emph{Primary Layout} with parallelization: $221.39$ ms, $242.56$ ms, $263.73$ ms with 1, 2 and 4 threads respectively. 

We compare the energy usage values of Servo in \emph{Overall Layout} using 1, 2, and 4 threads. Although the data corresponds to the processor energy usage between the beginning and termination of Servo, these values are mainly affected by parallelization of \emph{Overall Layout} because this stage constitutes the majority of the browser's execution time. We cannot obtain energy measurements at the granularity of function calls using Powermetrics~\cite{powermetrics}, or any external energy profiling tool.

\subsection{Performance Cost Model}
\label{sec:perfcostmodel}

In the Performance Cost Model, we consider only parallel runtimes to label the web pages appropriately. Consider an arbitrary number of thread configurations where each configuration uses $t$ threads. The values of $t$ are distinct. We first define the following terms.
\begin{itemize}
	\item $x_{t}$ -- time taken by $t$ threads
	\item $t_{\text{serial}} = 1$ (serial execution)
	\item $p_{t} = x_{t_{\text{serial}}} / x_t$ (speedup)
	\item $p_{t}^{\text{max}}$ -- maximum value of $p_{t}$
	\item $p_{\text{min}}$ -- minimum threshold that demarcates a significant speedup (to disregard measurement-noise)
\end{itemize}

The following steps describe the labeling process for a single web page:
\begin{enumerate}[1.]
	\item For each thread configuration, we compute its speedup with respect to serial execution.
	\item We calculate $p_{t}^{\text{max}}$ for a web page using a maximum operation on the set of all its $p_{t}$ values.
	\item If $p_{t}^{\text{max}} > p_{\text{min}}$, we assign a label $t$ where $t$ corresponds to that of $p_{t}^{\text{max}}$. Otherwise, we label the web page as $t_{\text{serial}}$ since all other $p_{t}$ values would be smaller than $p_{\text{min}}$.
\end{enumerate}

\begin{algorithm}[ht]
	\begin{algorithmic}[1]
		\State{{\bf Input:}}\\
		$T$: $\{\,t \mid t \text{ is number of threads in a configuration}\,\}$  \\
		$P$: $\{\,p_t \mid p_t \text{ is speedup using } t \text{ threads, } \forall\,t \in T\,\}$ \\
		$p_{\text{min}}$: minimum threshold for significant speedup 
		\Procedure{Performance-labeling}{}
		\State $p_{t}^{\text{max}} \leftarrow $ max$(P)$
		\If {$p_{t}^{\text{max}} > p_{\text{min}}$}
		\State $ label \leftarrow t$
		\Else
		\State $ label \leftarrow t_{\text{serial}}$
		\EndIf
		\State \Return label
		\EndProcedure
	\end{algorithmic}
	\caption{Performance labeling of a web page}
	\label{alg:classification-perf}
\end{algorithm}

Hence, if there are $n$ thread-configurations, we have $n$ possible labels. If the label of a web page is $t$, it means that using $t$ threads achieves the best performance for that web page. Note that these labels are \emph{nominal} values. They only identify the category and don't represent the total number of thread-configuration or their order. Algorithm~\ref{alg:classification-perf} formally describes the classification of web pages using the Performance Cost Model.

%

For our experimental testbed with 4 cores, we had three thread-configurations: 1 thread ($t = 1$), 2 threads ($t = 2$), and 4 threads ($t = 4$). For a browser, where the running times are in the order of milliseconds, even a small performance improvement is significant. However, to account for noise in our measurements, we consider a $10\%$ speedup to be significant. We attribute speedups less than 10\% to noise. Hence, we set the threshold value of $p_{\text{min}} = 1.1$. Using Algorithm~\ref{alg:classification-perf}, the total number of web pages categorized into labels $1$, $2$, and $4$ are $299$ $(55.88\%)$, $49$ $(9.15\%)$, and $187$ $(34.95\%)$ respectively.

\subsection{Energy Cost Model}
\label{sec:energycostmodel}

In the Energy Cost Model, we consider only the energy usage values to label the web pages. The algorithm for labeling is the same as Algorithm \ref{alg:classification-perf}. Instead of using speedup values, we consider greenup~\cite{choi2013} (energy usage improvement) values. Let $y_{t}$ represent the energy consumed by $t$ threads. For each thread configuration, we compute its greenup, $e_{t}$ with respect to serial execution ($e_{t} = y_{t_{\text{serial}}} / y_{t}$). $e_{\text{min}}$ is the minimum threshold that demarcates a significant greenup.

Since our experimental platform is a laptop, we did not use the Energy Cost Model to classify our web pages. We will consider this model in the future for energy-critical mobile devices.

\subsection{Performance and Energy Cost Model}
\label{sec:perfenergycostmodel}

In the Performance and Energy Cost Model, we consider both timing and energy usage values to label the web pages. In cases where we can guarantee significant performance improvements through parallelization, we also need to consider increases in energy usage. Spawning more threads could result in higher energy usage especially if the parallel work scheduler is a power-hungry one such as a work-stealing scheduler (as is the case currently in Servo). Hence, we consider performance improvements through parallelization to be useful only if the corresponding energy usage is lesser than an assigned upper limit. We label web pages using this cost model with a bucketing strategy as described below. 

Similar to the classification in the previous two cost models, we consider an arbitrary number of thread configurations where each configuration uses $t$ threads. Each thread configuration has a corresponding speedup, $p_t$ and a greenup, $e_t$. In addition to the terminology defined in the previous two subsections, we define the following terms.
\begin{itemize}
	\item $PET_t$ -- \emph{performance-energy tuple (PET)}, $\{\,p_t\text{, }e_t\,\}$ which represents the speedup and greenup achieved using $t$ threads.
	\item $P_jP_{j+1} $ -- \emph{PET bucket} to which a certain number of PETs belong. $P_j$ and $P_{j+1}$ represent speedup values where $j \in \mathbb{N}$. A PET, $PET_t \in P_jP_{j+1}$ if $P_j < p_t < P_{j+1}$. One can define an arbitrary number of such buckets to categorize the tuples. Note that the value of $P_1$ (lower limit of the first bucket) is always $p_{\text{min}}$.
	\item $E_j$ -- energy usage increase limit (defined in terms of greenup) for a performance bucket $P_jP_{j+1}$ where $j \in \mathbb{N}$. $E_j$ demarcates the tolerance of energy usage increase for all $PET_t \in P_jP_{j+1}$.
\end{itemize}

In this labeling, we perform the following steps for each web page:
\begin{enumerate}[1.]
	\item We ignore all the values of $p_t$ that are lower than $p_{\text{min}}$ and we define PET buckets based on design considerations.
	\item If the filtering results in an empty set, we label the web page as $t_{\text{serial}}$. Otherwise, we organize the remaining speedups and greenups into PETs and assign them to the right PET buckets.
	\item Starting from the last bucket (one with highest speedups),
	\begin{enumerate}
		\item We sort the PETs in the descending order w.r.t. $p_t$ values.
		\item We look at the PET with the highest speedup, $p_t$ within this bucket and check to see if the corresponding energy usage, $e_t$ is less than the bucket's energy usage limit, $E_j$. If the check is not satisfied, we look at the next largest speedup in this bucket and repeat this step.
		\item When all PETs in a bucket don't satisfy the condition, we look at a lower bucket (one with the next highest speedups) and repeat the process. We do so until a PET satisfies the check against the energy usage limit.
	\end{enumerate} 
	\item If none of the PETs satisfy the condition, we label the web page as $t_{\text{serial}}$. Otherwise, we label the web page as the value of $t$ corresponding to the first PET that satisfies the condition.
\end{enumerate}

\begin{figure*}[htbp]
	\begin{center}
		\includegraphics[width=\textwidth]{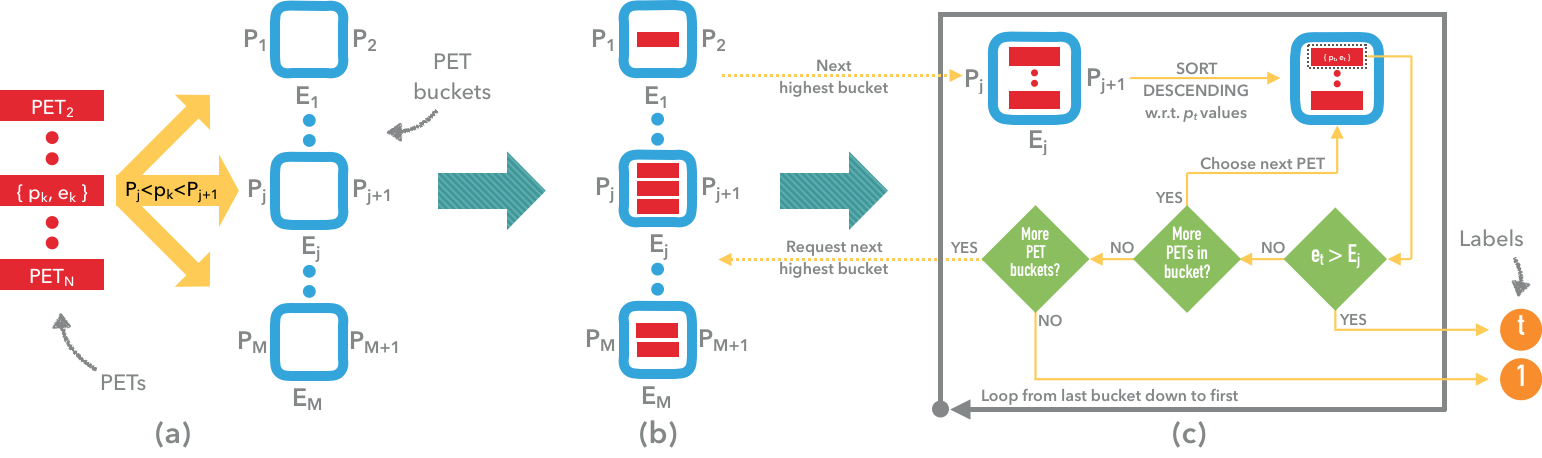}
	\end{center}%
	\vspace{-1em}
	\caption{A visual representation of the Performance and Energy Cost Model's labeling algorithm for a given web page. (a) $N$ PETs and $M$ PET buckets. (b) Based on $P_j$ and $P_{j+1}$ values, the PETs are assigned to the right buckets. (c) Algorithm-flow to choose the correct label for a web page.}
	\label{fig:petalgo}
	\vspace{-0.75em}
\end{figure*}

Algorithm \ref{alg:perf-energy-classification} formally describes the classification of the web pages using the Performance and Energy Cost Model and \figref{fig:petalgo} portrays a visual representation of the same.

\begin{figure}[htbp]
	\begin{center}
		\includegraphics[width=0.45\textwidth]{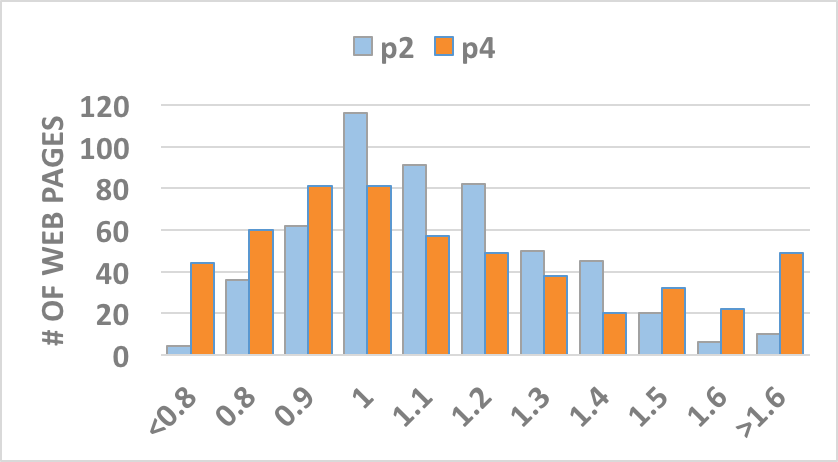}
	\end{center}%
	\vspace{-1em}
	\caption{Histogram of speedup values. The bin labels are upper limits.}
	\label{fig:hist}
	\vspace{-1em}
\end{figure}%
%

For our case study with Servo, we had three thread configurations: 1 thread ($t = 1$), 2 threads ($t = 2$), and 4 threads ($t = 4$). We set the value of $p_{\text{min}} = 1.1$ (from Section~\ref{sec:perfcostmodel}). \figref{fig:hist} depicts the histogram of Servo's $p_2$ and $p_4$ values. The histogram shows that, out of the significant speedups (\texttildelow 40\%), the half point lies roughly at $1.3$. Thus, we used two performance buckets: $P_1P_2$ and $P_2P_3$ where $P_1 = p_{\text{min}}$, $P_2 = 1.3$, and $P_3 = 12.87$ (the largest observed speedup). For the first bucket, we set the energy usage increase tolerance, $E_1 = 0.9$\footnote{These values can be tweaked based on design and device considerations. We choose these values for Servo on a quad-core Intel Ivy Bridge.} since a $10\%$ energy usage increase can make a noticeable difference in overall battery life of a laptop. For the second bucket, we chose a tolerance, $E_2 = 0.85$ since an energy usage increase beyond $15\%$ is not acceptable for any performance improvement. Using Algorithm \ref{alg:perf-energy-classification}, the total number of web pages categorized into labels $1$, $2$, and $4$ are $317$ $(59.25\%)$, $50$ $(9.34\%)$, and $168$ $(31.40\%)$ respectively.

\begin{algorithm}[ht]
	\begin{algorithmic}[1]
		\State{{\bf Input:}}\\
		$T$: $\{\,t \mid t \text{ is number of threads in a configuration}\,\}$  \\
		$PET$: $\{\,PET_t \mid PET_t=\{\,p_t, e_t\,\} \text{, where } \newline p_t \text{ is speedup using } t \text{ threads, } \newline e_t \text{ is greenup using } t \text{ threads, } \forall\,t \in T\,\}$  \\
		$P$: $\{\,P_jP_{j+1} \mid P_jP_{j+1} \text{ is a bucket of PETs whose } \newline P_j < p_t < P_{j+1}\,\}$ \\
		$E$: $\{\,E_j \mid E_j \text{ is energy usage increase limit for } P_jP_{j+1} \text{, } \linebreak \forall\, P_jP_{j+1} \in P\,\}$ \\
		$p_{\text{min}}$: minimum threshold for significant speedup  
		\Procedure{Performance-Energy-labeling}{}
		\State $label \leftarrow t_{\text{serial}}$
		\For {$P_jP_{j+1} \in P$} // highest $j$ to lowest $j$
		\State $PET^{'} \leftarrow \text{all } PET_t \in P_jP_{j+1}$
		\State $PET^{''} \leftarrow sortDescending(PET^{'}\text{ w.r.t }p_t)$
		\ForAll {$PET_t \in PET^{''}$}
		\If{$e_t > E_j$}
		\State $label \leftarrow t$
		\State break
		\EndIf
		\EndFor
		\EndFor
		\State \Return label
		\EndProcedure
		
	\end{algorithmic}
	\caption{Performance-Energy labeling of a web page}
	\label{alg:perf-energy-classification}
\end{algorithm}
\vspace{-0.5em}

\section{Performance Modeling and Prediction}
\label{sec:perf}

\begin{figure}[htbp]
	\begin{center}
		\includegraphics[width=0.49\textwidth]{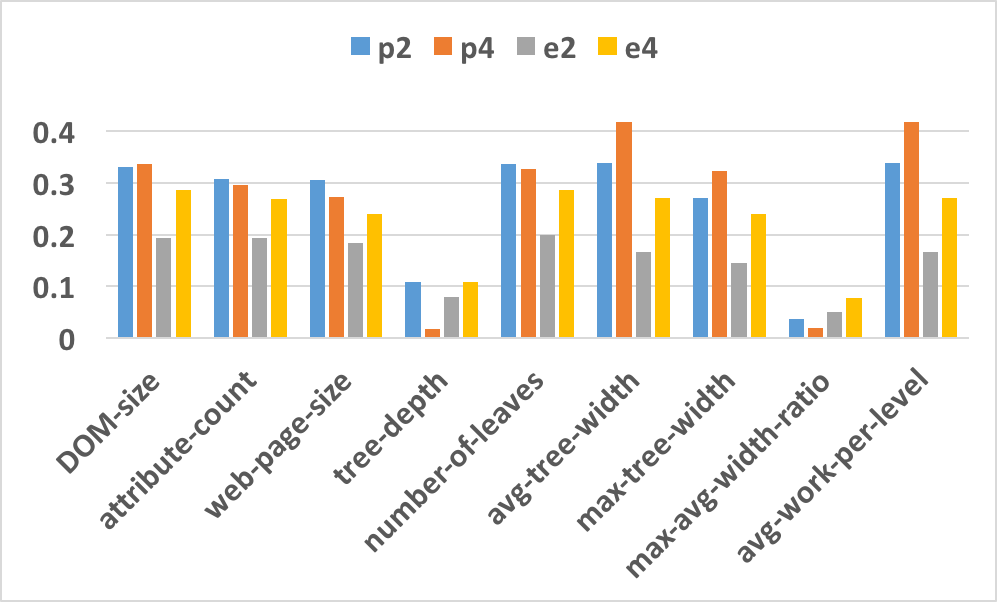}
	\end{center}%
	\vspace{-1em}
	\caption{Statistical correlation strengths: $R$-values.}
	\label{fig:corr}
	\vspace{-1em}
\end{figure}%

Our aim is to model the relationship between a web page's characteristics and the parallel performance of the web rendering engine to perform styling and layout on that page. With such a model, given a new web page, a browser will be able to predict the parallel performance improvement. The browser can then decide the number of threads to spawn during \emph{Overall Layout} for a given web page using a statistically constructed model. When parallelization is beneficial, the browser should also consider energy usage values and check for tolerable amounts. Hence, our goal is to build a predictive model that allows a browser to decide the number of threads to spawn for its \emph{Overall Layout} stage for any given web page by only looking at the web page's essential characteristics.

In this section, we construct and describe predictive models for Servo using off-the-shelf supervised learning methods on the Performance and Energy Cost Model (see Section~\ref{sec:perfenergycostmodel}). For each of the models, our predictive features are the characteristics of a web page that we describe in Section~\ref{sec:char}. Since the features lie on different scales\footnote{\eg{} \textbf{DOM-size}'s statistical range is 8307 while \textbf{avg-tree-width}'s is 436.}, we use Normal Standardization\footnote{The values of a feature are distributed according to its Normal distribution such that the mean of the feature is 0 and its standard deviation is 1.} (using MATLAB's \texttt{zscore}~\cite{zscore}) on the features to prevent one feature to dominate the other during learning. \figref{fig:corr} shows the correlation coefficient, $R$-values between the nine features and the speedups and greenups observed with 2 ($p_2$, $e_2$ values) and 4 ($p_4$, $e_4$ values) threads. We see that a positive linear\footnote{Statistical correlation measures the strength of only linear relationships between two variables.} relation exists between parallel benefits and the web page features. However, the exact relationship between the predictors and results is unlikely to be strictly linear, as is evident from our modeling results. Also, we see that \textbf{tree-depth} and \textbf{max-avg-width-ratio}, by themselves, do not correlate strongly ($<0.1$) with the speedups and greenups observed in our dataset. However, when used in combination with other features, as in \textbf{avg-work-per-level}, we note	a stronger relationship. For our modeling, we choose the seven predictor features that have $R$-values greater than $0.1$. The output for each of the models is one of the three labels: $1$, $2$, or $4$ representing 1 thread, 2 threads, and 4 threads respectively. These response labels are \emph{nominal} values. 

Our dataset has a \emph{large} class imbalance \ie{} one label has many more observations than the other. For the Performance and Energy Cost Model, only $49$ instances are labeled $2$ while $299$ and $187$ are labeled $1$ and $4$ respectively. Also, the predictor to observation ratio for our data set is quite small. Hence, we don't face the issue of over-fitting.

To validate and test our models, we use \emph{cross-validation} using a $90$-$10\%$ training-testing ratio \ie{} we divide our dataset into 10 subsets. Using each of these 10 subsets as the testing set ($55$ samples) and the remaining data as the training set ($480$ samples), we train 10 models. For each of the models, we predict labels for their corresponding testing sets, which results in 10 model prediction accuracies. We consider the mean and maximum of these 10 accuracies. When we are not able to use cross-validation, we use the \emph{holdout-set} technique (again using a $90$-$10\%$ training-testing ratio) wherein the data is divided into two sets: training and testing. We choose to use the cross-validation or holdout-set techniques because they result in true prediction error values and not ``model errors"~\cite{prederror}. Additionally, they don't rely on any parametric or theoretic assumptions about the data.

We experimented with the following supervised learning algorithms: Multinomial Logistic Regression (MNR), Ensemble Learning, and Neural Networks. We choose these three methods to capture non-linear relationships between web page characteristics and a rendering engine's parallel performance.


\textbf{MNR}: In this multinomial logit model, the probability of each label for a web page is expressed as a non-linear function, using logit link functions of the predictor tree characteristics. We trained our model using MATLAB's \texttt{mnrfit}~\cite{mnr} function. With a $90$-$10\%$ training-test divide for cross-validation, we observe a mean and maximum accuracy of \textbf{$72.22\%$} and $87.27\%$ respectively.

\textbf{Ensemble Learning}: For Ensemble Learning framework, the results of many models are combined to generate the final prediction. For our data, we used the AdaBoostM2 (a variant of Adaptive Boosting for multi-class data) learning method on $100$ simple Decision Tree learners and also on $100$ Decision Trees with surrogate splits. We did so using MATLAB's \texttt{fitensemble}~\cite{fitensemble} function. Using regular trees, with a $90$-$10\%$ divide, we observe a mean accuracy of $71.12\%$ and a maximum of $83.63\%$. Using trees with surrogate splits, we see a mean accuracy of \textbf{$69.44\%$} and a maximum of $85.45\%$.

\textbf{Neural Networks}. Artificial Neural Networks consist of simple, connected elements. By training the weights of the connections between different elements, a large neural network can capture complex relationships between variables. For our data, we use a small neural network with $1$ hidden layer containing $10$ neurons. We do so using MATLAB's \texttt{nprtool}~\cite{nprtool}. Since MATLAB currently does not support cross-validation for its classifying neural network models, we use the holdout-set method to measure the accuracy of this model. We use $80\%$ of the data for training, $10\%$ for validation (which is essentially a part of training), and $10\%$ for testing. The accuracy of this model on the testing set is $77.8\%$; the accuracy on all of the data is $71.61\%$. We attribute these lower (than those of MNR and Ensemble Learning) accuracies to the simplicity of the model and the lack of cross-validation.


The high accuracies of these off-the-shelf learning methods emphasize the effectiveness of our automated labeling algorithms. Instead of concentrating our efforts on tweaking the parameters of machine learning techniques to extract higher accuracies on our limited dataset, we demonstrate high accuracies by using \emph{accurate} labeling algorithms that can be used on any dataset.




\textbf{Limitations}. These accuracies, however, are not greater than $90\%$. Machine learning models have the potential to be highly accurate but are heavily dependent on a large amount of accurate training data. They behave as ``black boxes" with the actual underlying relationships between variables remaining undiscovered to the user~\cite{warwick2004, alpaydin2010, warwick2012}. Our data for this case study is relatively small and is from a prototype; Servo is a project under development and is undergoing constant change. Many components of Servo haven't been optimized for performance yet. Also, with multiple threads being spawned to exploit and explore concurrency in browser tasks, repeatability of executions is hard to acquire. Consequently, our data has a fair share of outliers---$35\%$ of the working set of web pages observes a MAD greater than $25$. The class imbalance in our dataset is also an important factor that influences model accuracies since we don't have an equal share of training data for each class. 


Despite our limitations, we observe \emph{confident} accuracies on each of these models. This exactitude fosters our intuition that the web page characteristics are indeed related to the parallel performance and energy usage of a browser and that the practical parallelism benefits are predictable by these features. The best performing model, MNR, compared to the current implementation of Servo, achieves a maximum of $94.52\%$ performance savings (2.48 ms with 1 thread vs. 45.41 ms with 4 threads on \texttt{indeed.com}) and a maximum of $46.32\%$ energy savings (84.88 J with 1 thread vs. 158.14 J with 4 threads on \texttt{starbucks.com}). By performance savings of $x\%$, we mean our model shaves off $x\%$ of the program's execution time. Similarly, by energy savings of $y\%$, we mean our model shaves off $y\%$ of the program's energy usage. 

\section{Related Work}
\label{sec:related}

Research on parallel browser architectures and browser tasks began only recently, starting, to the best of our knowledge, in 2009. Although multi-core processors are ubiquitous today on both laptops and mobile devices, the browsers are yet to utilize their benefits. The growing concern of slow page load times, especially on mobile devices, and the unexploited parallelism benefits in commodity browsers are the primary motivations for this ongoing research. Below we outline existing research on parallelizing and analyzing browser-tasks.

\textbf{Browser Workload Characterization}. Gutierrez et al.~\cite{gutierrez2011} present characterization of an Android browser at the micro-architecture level using $11$ web pages. Our approach, on the other hand, is agnostic to the platform on which a browser runs. Zhu et al.~\cite{zhu2013} correlate the web page variances to the difference in page load times and energy usage of the \emph{serial} Firefox browser by characterizing the HTML and CSS elements of a web page. Our approach is similar in spirit: we consider only web page features to be predictive of a browser's performance. However, we find additional DOM tree features that are representative of the degree of the \emph{parallel} workload in a page.

\textbf{Browser Performance-Energy Analyses}. Thiagarajan et al.~\cite{thiagarajan2012} present a breakdown of energy usage by the different elements, such as CSS and Javascript, of a serial browser. Additionally, they propose a few optimizations to improve the power consumption of web browsers, such as re-organizing Javascript files or removing unnecessary CSS rules. Zhu et al. also focus on scheduling methods of heterogeneous systems to improve the energy efficiency of mobile processors. They evaluate benefits of a big.LITTLE heterogeneous system for a trade-off between performance and energy. However, both these projects assess on serial browsers. Our work, on the other hand, analyzes performance and energy trade-offs for a \emph{parallel} browser while remaining agnostic to the browser implementation and execution platform. To the best of our knowledge, this work is the first of its kind.

\textbf{Parallel Browsers}. In general, browsers use processes to isolate tabs and windows to enhance security. Gazelle~\cite{wang2009multi} uses two processes per page while Chrome uses a process-per-page approach. This method doesn't exploit parallelism within the tasks. ZOOMM~\cite{ZOOMM} explores the challenges in managing concurrency in multi-core mobile-device browsers. This work exploits parallelism within the styling, image decoding, and JavaScript tasks leaving layout as future work. Mai et al. ~\cite{mai2012case} propose that browser developers should focus on parallelizing web pages rather than browser-tasks. They build Adrenaline, a server-client browser. On the other hand, Mozilla Research's Servo exploits both safety, particularly through its use of Rust, and parallelism between browser tasks and also within the tasks themselves. So far, Servo has exploited parallelism in its styling, layout and painting tasks, and is continuing to explore parallelism in other computational bottlenecks such as parsing. Servo is being developed for Android as well. More importantly, our work is not on parallelizing layout but is in predicting the degree of parallelism inherent in rendering a web page by considering the parallel performance and energy usage of a browser.  

\textbf{Parallelizing Browser Tasks}. Several research projects on parallelizing browser-tasks such as styling~\cite{meyerovich2010fast, badea2010towards} and parsing~\cite{hpar} exist. Meyerovich et al.~\cite{meyerovich2010fast} introduce fast and parallel algorithms for CSS selector matching, layout solving and font rendering, and demonstrate speedups as high as $80\times$ using $16$ threads for six websites. However, they implement only a subset of CSS (without cascading) and evaluate their algorithm in isolation (not within a browser) without considering effects on energy usage. Servo adheres to the complete CSS specification, and our work is modeled within the \emph{complete} execution of a browser while considering energy usage constraints.

\section{Conclusion}
\label{sec:conclusions}

The workload of a web rendering engine is dependent on the web page it is rendering; we model the relationship between key web page features and the parallel performance of the rendering engine using supervised learning methods. Specifically, we characterize web pages using DOM tree and HTML characteristics that correlate to the \emph{Overall Layout} task but are blind to the rendering engine's implementation. We propose accurate and tunable, automated labeling algorithms that categorize web pages into a user-defined number of classes. Moreover, our algorithm accounts for trade-offs between performance improvements and energy usage increases for multi-core processors. Using multinomial logit classification, ensemble learning, and simple neural networks, we demonstrate robust predictive model accuracies, achieving $87.27\%$ with 535 web pages within the complete execution of a browser. On a laptop platform, our best performing model delivers performance and energy savings up to $94.52\%$ and $46.32\%$ respectively.



\section*{Acknowledgments}
We would like to thank Sean McArthur from Mozilla, Subramanian Meenakshi Sundaram and Forough Arabshahi from the University of California, Irvine and others from Mozilla Research for helping us in conducting and analyzing our experiments. 

\bibliographystyle{abbrv}
\small
\bibliography{main}
\normalsize
\end{document}